\documentclass[conference]{IEEEtran}
\usepackage[utf8]{inputenc}
\usepackage{hyperref}
\usepackage{listings}
\usepackage{color}
\usepackage{subcaption}
\usepackage{graphicx}
\usepackage{soul}
\usepackage{todonotes}

\definecolor{javared}{rgb}{0.6,0,0} 
\definecolor{javagreen}{rgb}{0.25,0.5,0.35} 
\definecolor{javapurple}{rgb}{0.5,0,0.35} 
\definecolor{javadocblue}{rgb}{0.25,0.35,0.75} 

\newcommand{\rname}{SPOC}
\newcommand{\ie}{{\em i.e.,\/ }}
\newcommand{\eg}{{\em e.g.,\/ }}

\newcommand{\etal}{{\em et al.\/}}

\begin{document}
%
\title{SPOC: Secure Payments for Outsourced Computations}

\author{\IEEEauthorblockN{Michał Król}
\IEEEauthorblockA{University College London\\
Fluentic Networks, UK\\
m.krol@ucl.ac.uk}
\and
\IEEEauthorblockN{Ioannis Psaras}
\IEEEauthorblockA{University College London\\
Fluentic Networks, UK\\
i.psaras@ucl.ac.uk}
}


\maketitle

\begin{abstract}
Constrained devices in IoT networks often require to outsource resource-heavy computations or data processing tasks. Currently, most of those jobs are done in the centralised cloud. However, with rapidly increasing number of devices and amount of produced data, edge computing represents a much more efficient solution decreasing the cost, the delay and improves users' privacy. To enable wide deployment of execution nodes at the edge, the requesting devices require a way to pay for submitted tasks. We present \rname~- a secure payment system for networks where nodes distrust each other. \rname~allows any node to execute tasks, includes result verification and enforce users' proper behaviour without 3rd parties, replication or costly proofs of computations. We implement our system using Ethereum Smart Contracts and Intel SGX and present first evaluation proving its security and low usage cost. 
\end{abstract}

\section{Introduction}


The Internet of Things (IoT) holds tremendous potential to change many of our daily activities, routines and behaviors. The pervasive nature of the information sources means that a great amount of data pertaining to possibly every aspect of human activity, both public and private, will be produced, transmitted, collected, stored and processed. Already heavily deployed smart devices are expected to drastically rise in numbers in future years \cite{danova2013morgan} consequently increasing the volume of produced data. At the same time, constrained IoT devices do not have enough computational power, making local processing of such amounts of data impossible. 

In the current model, the data is being transmitted to the cloud for further processing and aggregation. However, there is a constant stream of evidence that the centralized, data-center model of cloud computing appears inferior in light of emerging applications. Data flow exceeding backbone capacities and demand for minimal latency, in some cases sub-$ms$ response times, is in fact excluding classic cloud-based solutions. Furthermore, not all data produced by IoT devices might be of further use, or need to be permanently stored in backend cloud environments (\ie only the result of a computation on a data set might be of further interest).
 
As a result, recent research efforts have focused on distributed cloud infrastructures, where requested applications (or functions) are executed in specialized execution nodes at the edge of the network  \cite{ahmed2016survey} \cite{krol2017nfaas}. Such an infrastructure can provide lower delay, minimize the traffic that needs to travel all the way to the back-end data center, and reduce the exposure to failures of a single cloud provider.

However, for such system to be deployed, an efficient and secure payment system is essential and can determine its future success. In an open, non-walled garden cloud computing environment, execution nodes are owned by multiple stakeholders, while requestors do not know which nodes will execute their tasks and thus whom to pay in advance. What is more, even with this knowledge they do not want to pay for yet unfinished or unverified tasks. On the other hand, an execution node receiving a request does not want to use its resources without making sure that it will eventually receive the corresponding payment.

To make a payment system truly distributed, one needs to include result verification techniques to ensure its correctness. However, already proposed schemes repeat the same computations on multiple nodes, produce costly, cryptographic proofs of computation\footnote{not to be confused with Proof of Work (PoW) used in Bitcoin \cite{nakamoto2008bitcoin}} or involve 3rd parties to act as a middleman (see Sec. \ref{sec:related_work}). All those techniques require requesting nodes to pay orders of magnitude more than the actual cost of the computations.

In this paper, we propose \rname~: Secure Payments for Outsources Computations. \rname is a distributed payment system, which allows the transfer of funds between mutually distrusting requesting and executing nodes. In \rname, requestors submit tasks to a blockchain and allow any node to claim it for execution. The blockchain does not belong to any central entity and its integrity is assured by thousands of miners charging only a minimal fee (see Sec. \ref{sec:blockchain}). When the computations are finished the result is returned to the requestor and the executing node is paid for its work. Our solution leverages deposits, smart contracts and \emph{trusted execution environments} (TEEs) to ensure proper behaviour of all parties involved without establishing any trust relation between them. \rname~does not involve a 3rd party to resolve conflicts, perform redundant computations  nor create costly cryptographic proofs. 


\section{Background}
\label{sec:background}
\subsection{Blockchain and Smart Contracts}
\label{sec:blockchain}
The blockchain technology \cite{nakamoto2008bitcoin} implements a distributed, append-only ledger in the form of connected blocks. Once information is stored in the blockchain it cannot be removed or altered. Network participants use a consensus protocol to agree on current state of the ledger. As long as the majority of participants are honest, the integrity of the ledger is assured. Blockchain is widely used to record transactions in multiple crypto-currencies (\ie Bitcoin \cite{nakamoto2008bitcoin}, Ethereum \cite{buterin2013ethereum}). A common extension consists of scripting languages enables to include logic as part of the transaction and allows deployment of Smart Contracts - code submitted to blockchain and executed by all miners. Solidity - the scripting language of Ethereum is Turing-complete allowing to implement wide range of applications running on the blockchain.

\subsection{Trusted Execution Environment}
\label{sec:tee}

Intel Secure Guard Extension (SGX) is an example of Trusted Execution Environment (TEE) that allows applications or part of an application code to be executed in a secure container, called \textit{enclave}, protecting the integrity and confidentiality of data. SGX protects data from other applications and privileged system software, such as the operating system (OS), hypervisor, and BIOS. SGX implements hardware encryption in the CPU. In order to achieve protection against privileged software, the memory content of the enclave is stored inside the Enclave Page Cache (EPC), which is protected memory where encrypted enclave pages and Intel SGX data structures are stored. A page table that maps enclave pages onto EPC frames is managed by the OS. However, the OS cannot see the memory content because the EPC region is encrypted by the Memory Encryption Engine (MEE) within the CPU. It is only the enclave that is associated with the EPC page that can access content within the enclave.

In order to enable an application to use enclaves, the developer must provide a signed shared library that will execute inside an enclave. The library itself is not encrypted and can be inspected before being started, hence no secret should be stored inside the code. However, the SGX environment also provides a remote attestation protocol\cite{intel_remote}. Using this mechanism a user can verify the identity and integrity of a target enclave running on a remote host, and securely transfer confidential data using TLS \cite{gueron2016memory}.

The SGX does not introduce significant overhead and does not increase the execution time \cite{zhao2016performance}. In its current version SGX is susceptible to cache-timing attacks \cite{Gotzfried:2017:CAI:3065913.3065915}, however an enhancement has already been proposed to eradicate this vulnerability \cite{shih2017t}.

\section{Threat Model and assumptions}

We consider that the following actors participate in the transaction:
\begin{itemize}
  \item{Requestors - constrained IoT devices submitting resource-heavy tasks to the network and making the payment.}
  \item{Executing Nodes - high-capacity nodes using their resources to execute requested tasks/functions and receiving the payment.}
  \item{Payment System - a smart contract running on a blockchain.}
\end{itemize}

Our threat model assumes that when the Requestor submits its task to the blockchain, it does not know which node will execute it. We assume that for the requestor, the value of the executed task ($V_{Ri}$) is higher than the price she is willing to pay ($P_{Ri}$). At the same time, the cost of running the computation ($C_i$) is lower than the offered price ($P_{Ri}$). Both parties wish to finalise the transaction, but mutually distrust one another. Each party is potentially malicious, \ie they may attempt to steal funds, avoid making payments, and forge results if it benefits them. Any time each party may drop, send, record, modify, and replay arbitrary messages in the protocol. We assume that each executing node has a TEE-capable machine, while requestors are resource-constrained devices capable of performing simple computations only. 

Both the requestor and the executing node trust the blockchain, their own environments, the TEE, and the function running in the enclave. The rest of the system, such as the network between the parties and the other party's software stacks (outside the TEE) and hardware are untrusted. During function execution, the execution node may therefore: (i) access or modify any data in its memory or stored on disk; (ii) view or modify its application code; and (iii) control any aspect of its OS and other privileged software.
Our threat model takes into account denial-of-service attacks, where an executing node claims a task and never executes it, delaying reception of valid results by the requestor. 
We model requestors and executing nodes as individual rational adversaries. Rational means that a party always acts in a way that maximizes its payoff, and is capable of thinking through all  possible outcomes and choosing strategies which will result in the best possible outcome. 

\begin{table}[h!]
\begin{center}
\begin{tabular}{ | c | c| } 
\hline
$EN$ & Execution Node\\ 
\hline
$R$ & Requestor\\ 
\hline
$T_i$ & task $i$\\
\hline
$V_{Ri}$ & value of tasks $i$ for Requestor\\
\hline
$P_{Ri}$ & Requestor's payment for task $i$\\ 
\hline
$D_{Ri}$ & Requestor's deposit for task $i$\\ 
\hline
$D_{Ei}$ & Executing Node's deposit for task $i$\\
\hline
$C_i$ & cost of executing task $i$\\
\hline
\end{tabular}
\end{center}
\vspace{-10pt}
\caption{Notation}
\label{tab:notation}
\vspace{-4pt}
\end{table}

\section{Smart Payments}
\rname~operates as follows. A requestor submits its task to the blockchain specifying the reward. Any execution node can then claim the task and start the computation. Once the task is successfully claimed it is assigned to the execution node and cannot be claimed by others. When the computation is finished, the results are sent to the requestor (or other specified node) and the execution node receives its payment. The proper behaviour of both parties is assured by deposits that are sent back after successfully finalising the transaction.

\subsection{System Overview}

\begin{figure*}[ht]
\begin{subfigure}[t]{.5\textwidth}
  \centering
  \includegraphics[width=1\textwidth]{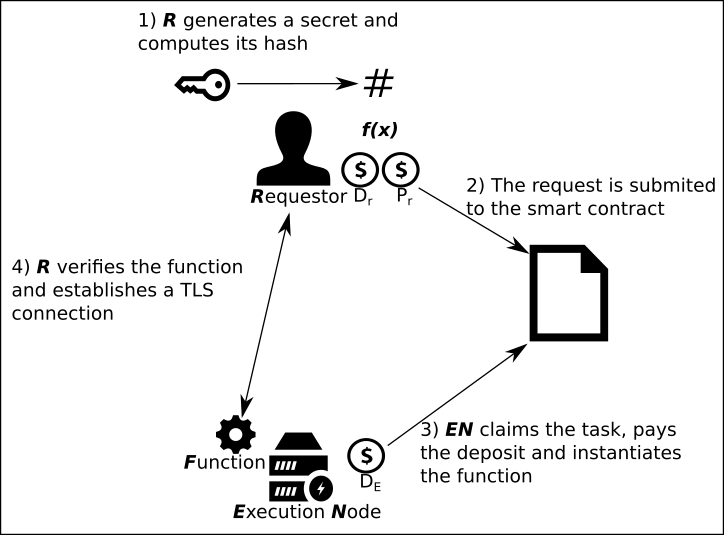}
\end{subfigure}
\begin{subfigure}[t]{.5\textwidth}
  \centering
  \includegraphics[width=1\textwidth]{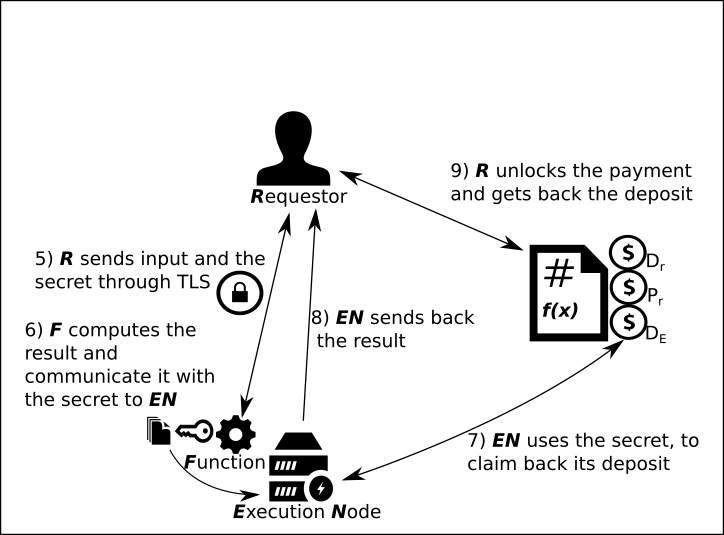}
\end{subfigure}
\vspace{-17pt}
\caption{A payment scheme using smart contracts.}
\label{fig:payments}
\vspace{-15pt}
\end{figure*}

In \rname, before submitting a task $T_i$, a Requestor (\textbf{R}) starts by generating a secret $S_i$  and computing its hash $H_i$ (Fig. \ref{fig:payments}). The task is then submitted to the smart contract. The submission includes a name of the function to execute, the hash, payment $P_R$ and deposit $D_R$.  All information stored in the contract is publicly visible, so the secret itself cannot be submitted. Both $P_R$ and $D_R$ are locked in the contract and cannot be manually retrieved by the requestor. $D_R$ prevents the requestor from receiving the result and not unlocking the payment for the node that did the computation. When an executing node (\textbf{EN}) wants to execute the task, it must invoke a claim function on the contract. The claim must contain another deposit $D_E$ that will be lost if the executing node does not execute the claimed task. As both $D_R$ and $D_E$ ensure proper behaviour of the requestor and the execution node respectively, they must be higher than a specified threshold. To prevent parallel execution, only the first claim with a sufficient deposit is accepted. The executing node must then instantiate the requested function. The function's authenticity is verified by the requestor using the Remote Attestation Protocol (Sec. \ref{sec:tee}). If the check is successful, the requestor establishes a secure TLS connection with the function and uses it to transmit input parameters and secret $S_i$. When the function finishes the computation, it returns the result and the secret to the executing node. The result is not sent back to the requestor using the TLS connection, as it would require the requestor to stay online. Also, in many scenarios the result must be reported to a third entity (\ie the cloud) and not to the requesting node. The executing node can now submit the received secret to the smart contract and receives back its deposit $D_E$. The result must be now sent to the indicated location (the requestor or a third party). When the requestor confirms a proper transmission of the result, it unlocks the payment on the smart contract. The executing node receives its payments $P_R$ and the requestor gets its deposit $D_R$ back. 

\rname~consists of 3 modules residing on the requestor, execution node and the blockchain, respectively.

\subsection{Requestor}
The module residing on the requestor resides on a constrained device and cannot execute resource heavy computations. The requestor must generate a short random secret and computes its hash. It then submits the task to the blockchain. Once the task is claimed, the requestor verifies the remote function and transmits the secret, the input parameters and the data encryption key through TLS to the executing node. The function must be trusted by the requestor and the function's behaviour must be verified in advance \cite{krol2017open}. Data produced by the execution node should be encrypted and must be signed to allow easy verification. If the check is successful, the requestor contacts the blockchain again to unlock the payment. 

\subsection{Execution Node}
We implement a simple client residing outside the TEE, that observes the blockchain. If a new task is submitted, the executing node tries to claim it and if successful (\ie no other node has requested it already), it downloads\footnote{Depending on the more general setup, the function might already be in local storage, \eg \cite{krol2017nfaas}} and instantiates the corresponding function. The function should be verified\footnote{\ie using techniques discussed in \cite{krol2017open}} before instantiating to avoid executing potentially malicious code. Each function creates its own enclave to securely store the secret, input parameters and produced results. The results can be either sent back using established TLS connection, or encrypted using a key provided by the requestor and sent using an open channel. The SGX system provides each enclave with a seal key that can be used to store data on stable storage and access it again upon subsequent execution. This facilitates the development of applications that can restart an enclave without requiring a new remote attestation in case of system failure. The enclave instead loads its secrets from a configuration file encrypted with the enclave specific seal key and kept in stable storage such as a hard drive. When a function finishes the computations, the result is sent to the specified destination and the whole enclave is removed together with produced temporary files. 

\subsection{Smart Contract}
We implement our smart contract in Solidity, the language used by Ethereum \cite{buterin2013ethereum}. The pseudo-code is presented in Lst. \ref{alg:contract}. The requestor starts by calling ``submitTask'' with the function name and a hash of the generated secret. The invocation must be combined with the transfer of funds containing both a payment and a deposit. If the submitted amount is not sufficient, the transaction will be rejected. A successful invocation creates a task that is stored on the blockchain. The execution node can then claim the task using the task ID and submitting a deposit. When the execution node receives the secret after finishing the computation, it can call ``\texttt{finalizeExecutionNode}'' to get back its deposit. The function compares the secret against its hash submitted by the requestor. The money transfer is made only if the check is successful. When the requestor receives the result, it calls ``\texttt{finalizeRequestor}''. The smart contract returns the deposit to the requestor, sends the payment to the execution node and deletes the task. Finally, we implement a timeout function. If the task is not executed within a specified threshold, the requestor can use it to get back the allocated payment. Both deposits remain locked in the contract and cannot be used anymore. The threshold value can be set depending on the application and is publicly visible for potential execution nodes. We also implement a set of contract events allowing interested parties to be notified when a new task is submitted, claimed, finished or times out.

\lstset{language=Java,
basicstyle=\tiny,
keywordstyle=\color{javapurple}\bfseries,
stringstyle=\color{javared},
commentstyle=\color{javagreen},
morecomment=[s][\color{javadocblue}]{/**}{*/},
stepnumber=2,
numbersep=8pt,
tabsize=4,
showspaces=false,
showstringspaces=false,
caption={\rname~contract.},
label=alg:contract}

\begin{lstlisting}
contract SmartPayments {
  
    function submitTask(string functionName, 
			bytes32 hash,
			uint expires) 
			public payable returns (uint taskID) {
        if(msg.value < THRESHOLD){
            msg.sender.transfer(msg.value);
            return;
        }
        taskID = numTasks++;
        tasks[taskID] = Task(functionName, 
			hash, 
			msg.sender, 
			msg.value - THRESHOLD, 
			THRESHOLD, 
			0x0, 
			0, 
			false, 
			false,
			now,
			expires);
        return taskID;
          
    }
    
    function claimTask(uint taskID) 
	      public payable returns (bool result) {
        if((msg.value < THRESHOLD) || 
	    tasks[taskID].claimed || 
	    tasks[taskID].completed){
            msg.sender.transfer(msg.value);
            return false;
        }
        
        tasks[taskID].executionNode_ = msg.sender;
        tasks[taskID].executionNodeDeposit_ = msg.value;
        tasks[taskID].claimed = true;   

        return true;
    } 
    
    
    function finalizeExecutionNode(uint taskID, bytes32 secret) 
	public returns (bool result){
        if((tasks[taskID].executionNode_ != msg.sender) || 
	  (!tasks[taskID].claimed) || 
	  (tasks[taskID].completed) || 
	  (sha256(secret) != secret)){
            return false;
        }
        tasks[taskID].completed = true;
        msg.sender.transfer(tasks[taskID].executionNodeDeposit_);
    }
    
    function finalizeRequestor(uint taskID) 
	 public returns (bool result){
         if((tasks[taskID].requestor_ != msg.sender) || 
	    (!tasks[taskID].claimed) || 
	    (!tasks[taskID].completed)){
             return false;
         }
         
         msg.sender.transfer(tasks[taskID].requestorsDeposit_);
         tasks[taskID].executionNode_.transfer(tasks[taskID].payment_);
         delete tasks[taskID];
    }
    
    function timeout(uint taskID) public returns (bool result){
        if(now > tasks[taskID].start + tasks[taskID].expires 
	  || !task[taskID].completed){
            tasks[taskID].requestor.transfer(payment);
            return strue;
        }
        return false;
    }
    
}
\end{lstlisting}

\section{Evaluation}
We evaluate \rname~ by performing a security analysis and running and investigating the behaviour of the smart contract on a test Ethereum network. We implement our solution using cpp-ethereum - a C++ Ethereum client\footnote{\url{https://github.com/ethereum/cpp-ethereum}} on a Dell XPS 13 laptop. The enclave runs with Intel SGX SDK for Linux v2.0. The performance of the execution node depends heavily on the submitted task. However, Intel SGX is proven to introduce only minimal overhead to the computations \cite{zhao2016performance}. 

\subsection{Security Analysis}
In this section, we provide the intuition behind the security properties of the protocol. We defer formal proofs of security to the extended version of the paper. 
Before submitting a task, the Requestor must generate a new secret and compute its hash. The hash is submitted to the blockchain and is thus publicly visible. We allow only one execution node to start working on the task and receive the allocated payment when finished. This sequence of events prevents parallel computations and protects from malicious users who could try to compute the secret from the hash and steal the funds. It also prevents the requestor (as he knows the secret) from stealing the payment from an execution node that is working on the task. The hash function does not have to be fully secure, but should be significantly more costly to reverse than computing the actual task itself. Even with fixed 32 byte input, calculating $2^{256}$ signatures is impossible within short amount of time. When a task times out and the requestor decides to resubmit it, the requestor must generate a new secret and hash. It is important to tune the ``expires'' parameter, so that the execution node has enough time to complete the task, but reversing the hash function should be impossible within this time window. 
The secret itself is known only to the requestor and transmitted through a secure channel to a protected enclave. The same applies to the submitted input. When the execution node receives the secret, it is submitted to the blockchain through an open connection and becomes publicly visible. However, the secret is useless for nodes that did not claim the task and can be used only by the valid node. The output data is encrypted and signed by keys provided by the requestors. Its integrity can then be easily verified and decrypted by authorised users. The executing nodes never hold the encryption key, it thus cannot forge results.

We investigate possible scenarios where both Requestor and Executing Node try to cheat and analyse their outcome (Tab. \ref{tab:outcome}). 
\begin{itemize}
\item \textbf{All parties behave honestly and the transaction is finalised} -  the Executing Node uses its resources to perform the computation ($C_i$. The Requestor receives the results ($V_{Ri}$) and the payment ($P_i$) is transferred to the Executing Node and both deposits are sent back to their owners. 
\item \textbf{An executing node claims a task, but does not execute it} - the requestor withdraws the task after a time out, getting back its money ($P_i$). Both deposits ($D_{Ri}$ and $D_{Ei}$) remain locked on the contract. This is because both $EN$ can decide not to compute the task, but $R$ can also prevent it by not sending input parameters and the smart contract is not able to distinguish between both cases. 
\item \textbf{An executing node computes the task, but does not send back the result} - the Executing Node is able to get its deposit back ($D_{Ei}$), but is not getting paid for performed computations, while using its resources ($C_i$) to compute the result. However, the requestor is not able to retrieve its deposit ($D_{Ri}$) nor the payment either ($P_i$). 
\item \textbf{A malicious requestor receives valid results, but does not confirm} - the malicious requestor loses it deposit ($D_{Ri}$), while the executing node is not getting paid for used resources ($C_i$).
\end{itemize}

Only when both parties behave honestly, they both gain value. In other scenarios, the cheating entity always loses funds. Unfortunately, removing any dispute resolution entity comes at a cost: an honest party can still lose money if its peer behaves improperly. In \rname, a rational attacker will always behave honestly, however a malicious adversary could cheat to make the other party lose funds. The smart contract can automatically unlock the funds if it could verify that the result was made available to the requestor. A naive solution consists of submitting results directly to the blockchain. However, storing large data sets significantly increase the cost of the transaction. A more viable approach involves distributed peer-to-peer storage such as IPFS \cite{benet2014ipfs}. The results must be made public in an encrypted and signed form, so that everyone, including the smart contract, can verify its authenticity, but only the requestor can decrypt the content.

\begin{table}[h!]
\begin{center}
\begin{tabular}{ | l | c | c | }
\hline
\textbf{Scenario} & \textbf{Requestor} & \textbf{Execution Node}\\ 
\hline
Both honest & $V_{Ri} - P_i$ & $P_i - C_i$ \\
EN does not execute & $-D_{Ri}$ & $-D_{Ei}$\\
EN does not sent the result & $-(P_i + D_{Ri})$ & $-C_i$\\
R does not confirm & $V_{Ri} - P_i - D_{Ri}$  & $-C_i$ \\
\hline
\end{tabular}
\end{center}
\vspace{-10pt}
\caption{Output of different system users' behaviours.}
\label{tab:outcome}
\vspace{-15pt}
\end{table}

\subsection{Smart Contract evaluation}
We analyse the cost of invoking each function of the contract and its deployment (Tab. \ref{tab:cost}). The tests were performed on test Rinkeby network\footnote{\url{www.rinkeby.io}}. 
Ethereum allows to specify a priority of newly submitted transactions. The slowest ones are processed within 10 minutes, standard within 5 minutes and the fastest within 2 minutes. With faster processing time comes increased transaction cost. With the slowest transaction processing, the system is cheap to exploit, keeping the cost much below 1\$\footnote{Costs calculated using \url{https://ethgasstation.info/}} (to be further split between requestors and execution nodes). However, with the fastest transaction, the cost increases up to almost 8\$ that is be unacceptable for smaller tasks. At the same time, for all the tested setups the delay is significant. The functions must be called sequentially and only if the previous call is confirmed to avoid double spending attacks \cite{karame2012double} (\ie an execution node must be sure that the task is submitted before claiming it). This means that the system is not suitable for delay-sensitive tasks, but this limitation is implied by the underlying blockchain technology (in our case - Ethereum). Involving a more lightweight distributed ledger \cite{al2017chainspace} could potentially overcome this limitation.


\begin{table*}[h!]
\begin{center}
\begin{tabular}{ | c | c | c| c | c | }
\hline
\textbf{Function} & \textbf{Gas} & \textbf{Ether Slow (\$)} & \textbf{Ether Standard (\$)} & \textbf{Ether Fast (\$)}\\ 
\hline
Deploy & 1260850 & 0.00013 (0.057\$) & 0.01387 (6.324\$) & 0.03656 (16.673\$)\\ 
\hline
submitTask & 277880 & 0.00003 (0.013\$) & 0.00278 (1.267\$) & 0.00806 (3.675\$)\\
\hline
claimTask & 145120 & 0.00001 (0.007\$) & 0.00145 (0.662\$) & 0.00421 (1.919\$)\\
\hline
finalizeExecutionNode & 52802 & 0.00001 (0.002\$) & 0.00053 (0.241\$) & 0.00153 (0.698\$)\\
\hline
finalizeRequestor & 106357 & 0.00001 (0.005\$) & 0.00117 (0.533\$) & 0.00308 (1.406\$)\\
\hline
\hline
Total per task & 582159 & 0.00006 (0.027\$) & 0.0064 (2.92\$) & 0.01688 (7.698\$)\\
\hline
\end{tabular}
\end{center}
\vspace{-10pt}
\caption{Cost of invoking contract functions.}
\label{tab:cost}
\vspace{-15pt}
\end{table*}

\section{Related Work}
\label{sec:related_work}
Multiple papers discuss result verification techniques. The first group focuses on constructing cryptographic proof of computation \cite{walfish2015verifying, oliaiy2017verifiable, fiore2016hash, wahby2016verifiable}. Such proofs are easy to verify without the need to re-execute the computations. However, the overhead of pre-computation and creation of the proof is orders of magnitude higher than the actual cost of the computation being verified.
Another approach consists of running the same computations on multiple servers \cite{canetti2011practical, distler2016resource, van2014versum, szajda2003hardening}. As long as a given fraction of servers is honest, the result can be guaranteed by a consensus protocol. These techniques require at least one honest server and significantly increase the overhead by repeating the computations. 

An approach closer to our work involving blockchain technology is \cite{dong2017betrayal}. The authors assume only two execution entities and design smart contracts discouraging them from colluding. Huang \etal~\cite{huang2016bitcoin} also distribute the task to multiple workers and exploit Commitment-based sampling \cite{du2010uncheatable} to verify the correctness of the result. Before starting the computations, the workers have to commit to the task by spending bitcoins that are lost in case of dishonest behaviour. However, both systems (\cite{dong2017betrayal, huang2016bitcoin}) still require to repeat computations and assume a trusted 3rd party to resolve conflicts. 

Multiple projects focus on incentivising fairness and timely delivery of the results using cryptocurrencies \cite{andrychowicz2014secure, bentov2014use, kumaresan2014use, kumaresan2016amortizing, chen2012efficient}. Workers deposit predefined amount of money that is lost if they misbehave. However, all of them focus on fairness rather than verifying produced results. 
Finally, several projects aim to facilitate blockchain-based micro-payments \cite{lind1612teechan, lundqvist2017thing, poon2015bitcoin, raiden}. Those projects are complementary to ours, and could be used to lower the cost and overhead of transactions.

%
%
%

\section{Conclusion}
We presented \rname, a new execution payment system for untrusted IoT networks. Our system allows any node to submit and execute tasks and enforces users' proper behaviour using Trusted Execution Environment (TEE) and Smart Contracts. Our solution is secure against rational adversary, cost efficient and introduces minimal computational overhead on IoT devices. 
In our future work, we plan to extend our system by a distributed results submission platform, making it fully resistant to arbitrary adversary as well. We also plan to investigate more lightweight distributed ledger solutions to decrease the exploitation cost and make the system suitable for delay-sensitive tasks. 

\section*{Acknowledgment}
This work has been supported by the EC H2020 UMOBILE project (GA no. 645124), the EC H2020 ICN2020 project (GA no. 723014) and the EPSRC INSP fellowship project (EP/M003787/1). 

\bibliography{biblio} 
\bibliographystyle{ieeetr}

\end{document}